\documentclass[twocolumn]{article}

\usepackage[numbers,comma,sort&compress]{natbib} 

\usepackage{filecontents}

\usepackage{times}
\usepackage[utf8]{inputenc}
\usepackage{microtype}
\usepackage{color}
\usepackage{graphicx}
\usepackage{booktabs} 
\usepackage{svg}
\usepackage{amsmath}
\usepackage{amssymb}
\usepackage[hidelinks]{hyperref}
\usepackage{xr}
\usepackage{calculator}
\usepackage{xcolor}

\definecolor{sergik}{rgb}{.6,0,.6}
\definecolor{hartmut}{rgb}{0,.6,.6}
\definecolor{oliver}{rgb}{.6,0,0}

\usepackage{etoolbox}

\makeatletter
\apptocmd{\thebibliography}{\global\c@NAT@ctr 0\relax}{}{}
\makeatother

\title{Accurate Machine Learned Quantum-Mechanical Force Fields for Biomolecular Simulations}

\author{
Oliver T. Unke$^{1,2,3\ast}$
\and
Martin St\"ohr$^{4}$
\and
Stefan Ganscha$^{1}$
\and
Thomas Unterthiner$^{1}$
\and
Hartmut Maennel$^{1}$
\and
Sergii Kashubin$^{1}$
\and
Daniel Ahlin$^{1}$
\and
Michael Gastegger$^{2,3,5}$
\and
Leonardo Medrano Sandonas$^{4}$
\and
Alexandre Tkatchenko$^{4\ast}$ 
\and
Klaus-Robert M\"uller$^{1,2,6,7,8\ast}$}

\date{
\normalsize{$^{1}$Google Research, Brain team}\\
\normalsize{$^{2}$Machine Learning Group, Technische Universit\"at Berlin, 10587 Berlin, Germany}\\
\normalsize{$^{3}$DFG Cluster of Excellence ``Unifying Systems in Catalysis'' (UniSysCat), Technische Universit\"at Berlin, 10623 Berlin, Germany}\\
\normalsize{$^{4}$Department of Physics and Materials Science, University of Luxembourg, L-1511 Luxembourg City, Luxembourg}\\
\normalsize{$^{5}$BASLEARN -- TU Berlin/BASF Joint Lab for Machine Learning, Technische Universit\"at Berlin, 10587 Berlin, Germany}\\
\normalsize{$^{6}$Department of Artificial Intelligence, Korea University, Anam-dong, Seongbuk-gu, Seoul 02841, Korea}\\
\normalsize{$^{7}$Max Planck Institute for Informatics, Stuhlsatzenhausweg, 66123 Saarbr\"ucken, Germany}\\
\normalsize{$^{8}$BIFOLD -- Berlin Institute for the Foundations of Learning and Data, Berlin, Germany}\\
\normalsize{$^\ast$To whom correspondence should be addressed}\\ \normalsize{E-mail:  oliver.unke@gmail.com, alexandre.tkatchenko@uni.lu, klaus-robert.mueller@tu-berlin.de}
}

\begin{document}



\twocolumn[
\begin{@twocolumnfalse}
	\maketitle
	\begin{abstract}
\noindent
Molecular dynamics (MD) simulations allow atomistic insights into chemical and biological processes.
Accurate MD simulations require computationally demanding quantum-mechanical calculations, being practically limited to short timescales and few atoms. For larger systems, efficient, but much less reliable empirical force fields are used. Recently, machine learned force fields (MLFFs) emerged as an alternative means to execute MD simulations, offering similar accuracy as \textit{ab initio} methods at orders-of-magnitude speedup. Until now, MLFFs mainly capture short-range interactions in small molecules or periodic materials, due to the increased complexity of constructing models and obtaining reliable reference data for large molecules, where long-ranged many-body effects become important.
This work proposes a general approach to constructing accurate MLFFs for large-scale molecular simulations (GEMS) by training on ``bottom-up'' and ``top-down'' molecular fragments of varying size, from which the relevant physicochemical interactions can be learned. GEMS is applied to study the dynamics of alanine-based peptides and the 46-residue protein crambin in aqueous solution, allowing nanosecond-scale MD simulations of $>$25k atoms at essentially \textit{ab initio} quality. Our findings suggest that structural motifs in peptides and proteins are more flexible than previously thought, indicating that simulations at \textit{ab initio} accuracy might be necessary to understand dynamic biomolecular processes such as protein (mis)folding, drug--protein binding, or allosteric regulation.
	\end{abstract}
\vspace{\baselineskip}
\end{@twocolumnfalse}
]

\clearpage
\section*{Introduction}

\begin{figure*}[!htb]
	\centering
	\includegraphics[]{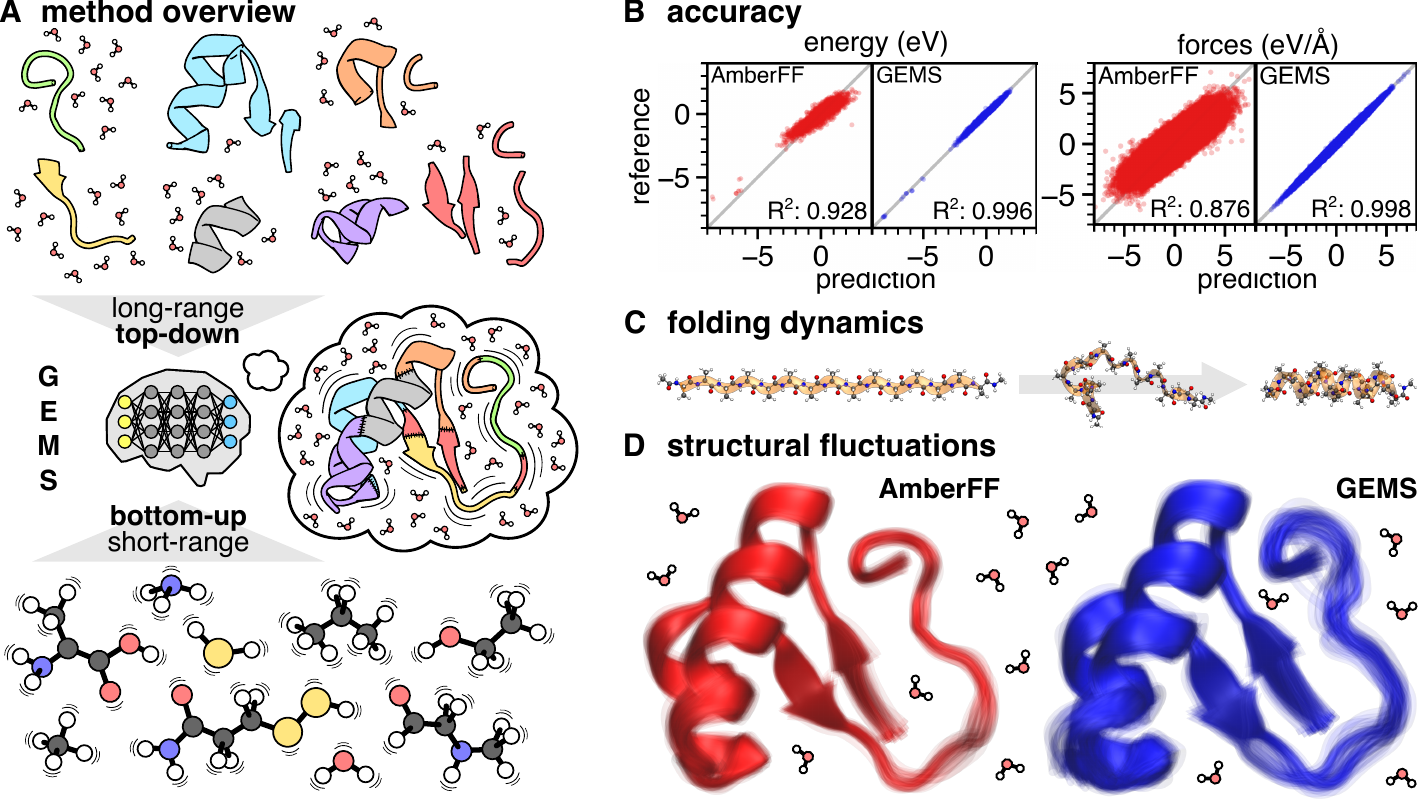}
	\caption{\textbf{Insights from GEMS simulations.} (\textbf{A}) Overview of the GEMS method. Different interaction scales on the potential-energy surface (PES) of a large system are learned from a combination of \textit{ab initio} reference data for top-down and bottom-up fragments. The resulting model is able to accurately reconstruct the PES of the molecular system and then used to study its dynamics.
		(\textbf{B}) Prediction accuracy for energies and forces of AceAla$_{15}$Nme conformations of GEMS compared to AmberFF \cite{Lindorff-Larsen2010AMBER99SB-ILDN} with respect to the PBE0/def2-TZVPP+MBD \cite{adamo1999toward,weigend2005balanced,tkatchenko2012accurate,ambrosetti2014} reference.
		(\textbf{C}) GEMS simulations show that the folding of AceAla$_{15}$Nme from a fully extended structure (left) to a helical conformation (right) occurs via intermediate conformations characterized by hydrogen bonding between backbone atoms of adjacent residues (middle). (\textbf{D}) Overlay of representative conformations (obtained from cluster analysis) sampled during an aggregated 10~ns of dynamics of crambin in aqueous solution. Simulations with GEMS (blue) lead to greater structural fluctuations compared to AmberFF (red), indicating that the protein is more flexible.}
	\label{fig:overview}
\end{figure*}

Molecular dynamics (MD) simulations allow to determine the motion of individual atoms in chemical and biological processes, enabling mechanistic insights into molecular properties and functions, as well as providing a detailed interpretation of experimental studies. MD simulations require a reliable model of the forces acting on each atom at every time step of the dynamics \cite{karplus2002molecular}. It is most desirable to obtain atomic forces from accurate solutions to the many-body Schr\"odinger equation \cite{schrodinger1926undulatory}, but this is only feasible for short MD simulations of few atoms for the foreseeable future \cite{mouvet2022recent}.

For larger systems, it is common practice to derive the forces from empirical models of the potential energy. Such force fields (FFs) approximate chemical interactions with computationally efficient terms and enable MD simulations of millions of atoms \cite{schulz2009scaling} for up to several milliseconds of dynamics \cite{shaw2008anton}. A disadvantage of FFs is their limited accuracy due to the neglect of important quantum-mechanical effects, such as changes to hybridization states, interactions between orbitals delocalized over several atoms, or electronic correlations between distant molecular fragments. Further, many FFs require a predetermined covalent bonding structure, preventing bond breaking and formation. When additional accuracy and flexibility is required, for example to study an enzymatic reaction, a possible alternative are quantum mechanics/molecular mechanics (QM/MM) simulations \cite{senn2009qm,mouvet2022recent}: The system is divided into a small QM region modelled with \textit{ab initio} methods (e.g.\ substrate and active site of an enzyme) and an MM region (e.g.\ the remaining protein and solvent molecules) described with an FF. However, the high computational cost associated with an accurate treatment of the QM region and the fact that it is often unclear which atoms need to be included for an adequate description of the process of interest \cite{kulik2016large} may limit the applicability of QM/MM methods.

In recent years, machine learned force fields (MLFFs) have emerged as an alternative means to execute MD simulations, combining the computational efficiency of traditional FFs with the high accuracy of quantum-chemistry methods \cite{unke2021machine}. To construct an MLFF, a machine learning (ML) model is trained on \textit{ab initio} reference data to predict energies and forces from atomic positions -- without the need to explicitly solve the Schr\"odinger equation outside of the reference data. MLFFs have led to numerous insights, e.g.\ regarding reaction mechanisms \cite{rivero2019reactive}, or the importance of quantum-mechanical effects for the dynamics of molecules \cite{sauceda2019molecular}. Despite these successes, until now, MLFFs have been applied primarily to MD simulations of small to medium-sized systems (tens of atoms), or periodic materials (e.g.\ metallic copper) \cite{jia2020pushing}. Applications to large heterogeneous systems, like proteins or other biologically relevant systems, have remained elusive, due to the increased complexity of constructing physically-informed ML architectures and obtaining reliable reference data for long-range interactions, which are known to play a key role in biomolecular dynamics~ \cite{rossi2015stability,stoehr2019quantum}.

This work proposes a general approach to constructing accurate machine-learned force fields for large-scale molecular simulations (GEMS). Based on the divide-and-conquer principle, MLFFs for large heterogeneous systems are trained on molecular fragments of varying size, which cover the relevant chemical interactions, but are still amenable to electronic-structure calculations. From this fragment data, the model infers to recompose the original system, which allows GEMS to successfully address the long-standing challenge of biomolecular simulations at \textit{ab initio} quality (Fig.~\ref{fig:overview}A). While MLFFs can successfully learn local chemical interactions from small molecules \cite{huang2020quantum}, a sufficient number of larger fragments is needed to learn long-range effects necessary to generalize to larger systems and achieve high prediction accuracy ($0.450$~meV/atom for energies and $36.704$~meV/\AA\ for forces). In this work, we rely on the recently proposed SpookyNet architecture \cite{unke2021spookynet}, which models dispersion and electrostatics explicitly by embedding physically motivated interaction terms into the ML architecture and learning their parameters from reference data. In addition, an empirical term for short-ranged repulsion between atomic nuclei increases the robustness of the model for strong bond distortions. SpookyNet also includes a mechanism to describe effects like non-local charge transfer, which other MLFFs (with some exceptions \cite{ko2021fourth}) are typically unable to. Taken together, these components enable the model to generalize to larger molecules when trained on appropriate reference data. 
However, ultimately, the quality and reliability of an MLFF should be judged by its predictions of experimental measurements -- for example, we show that GEMS is able to quantitatively reproduce experimental results regarding the helix stability of polyalanine systems at different temperatures and correctly describe infrared peak positions of water for a solvated protein.

GEMS is applied to MD simulations of model peptides and the 46-residue protein \textit{crambin} in aqueous solution ($>$25k atoms). When comparing to conventional force fields, such as AMBER99SB-ILDN \cite{Lindorff-Larsen2010AMBER99SB-ILDN} (AmberFF), GEMS approximates energies and forces computed from density-functional theory much more closely (Fig.~\ref{fig:overview}B). Interestingly, our findings reveal previously unknown intermediates in the folding pathway of poly-alanine peptides (Fig.~\ref{fig:overview}C) and a dynamical equilibrium between $\alpha$- and $3_{10}$-helices. In the simulations of solvated crambin, GEMS indicates that protein motions are qualitatively different and more flexible when compared to computations with a conventional FF (Fig.~\ref{fig:overview}D), showing contrasting short and long timescale dynamics. These results suggest that simulations at \textit{ab initio} accuracy may be necessary to fully understand dynamic processes like protein (mis)folding, drug--protein binding, or allosteric regulation.

\section*{Results}
\subsection*{Machine learning force fields for large systems}

\begin{figure*}[!htb]
	\centering
	\includegraphics[width=4.75in]{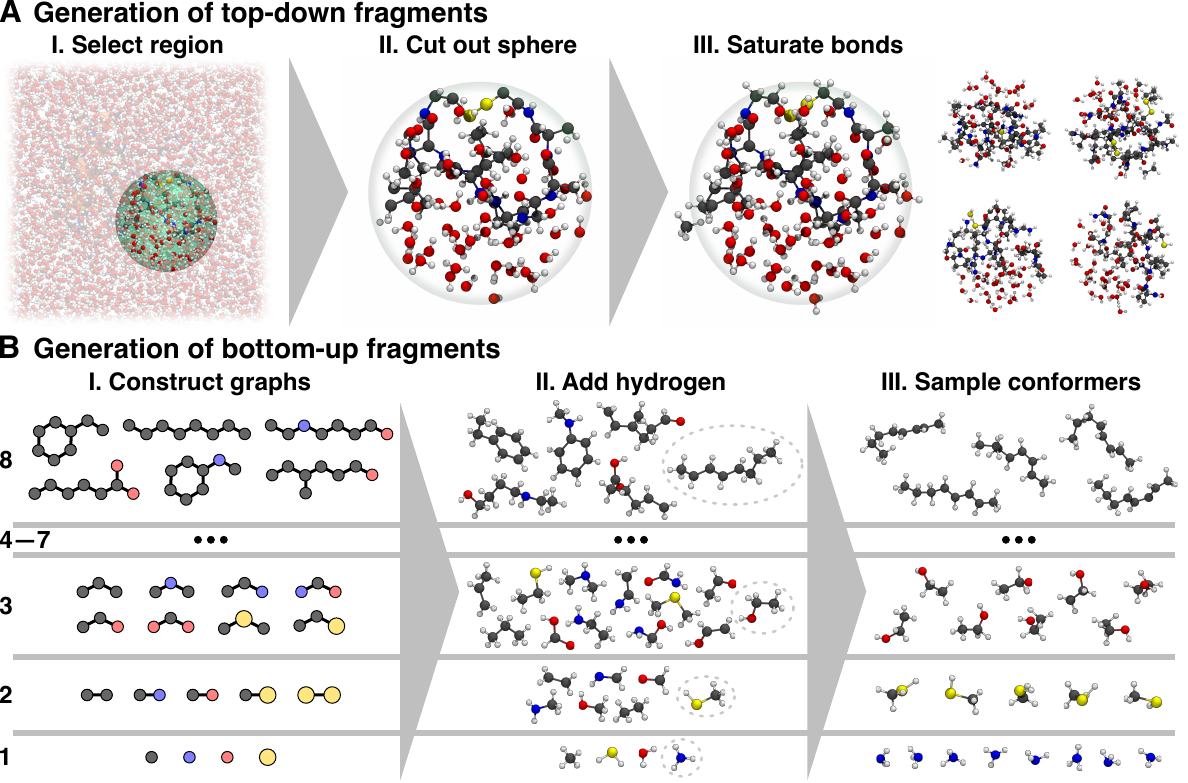}
	\caption{\textbf{Generation of top-down and bottom-up fragments.} (\textbf{A}) Top-down fragments are generated by cutting out a spherical region around an atom (including solvent molecules) and saturating all dangling bonds (the right side shows four top-down fragments generated from different regions). They are crucial for learning weak but long-ranged interactions, which are important for the dynamics of large systems. (\textbf{B}) Bottom-up fragments are generated by constructing chemical graphs consisting of one to eight non-hydrogen atoms (not all possible graphs are shown). The graphs are then converted to three-dimensional structures by adding hydrogen atoms. Due to their small size, multiple \textit{ab initio} calculations for many different conformers of each generated structure can be performed, allowing extensive sampling of the potential energy surface which is necessary for training robust models.}
	\label{fig:fragment_construction}
\end{figure*}

We start by generating reference data for smaller molecular fragments in order to train an MLFF, where the learned model accurately reflects the full large system. 
There are several strategies to achieve this goal. On one hand, the model needs to be able to learn all relevant chemical interactions that are necessary to reconstruct a complete and accurate picture of the system of interest from the fragment data. This is important to capture weak, but long-ranged interactions, which collectively dominate e.g.\ relative energy differences of different conformations of large molecules. On the other hand, it is necessary to prevent ``holes'' in the potential energy surface (PES) \cite{behler2014representing} -- regions with low potential energy corresponding to unphysical structures, e.g.\ featuring unnaturally large or short bond lengths. The existence of holes in the PES prevents stable MD simulations, because long trajectories eventually may become trapped by such artefacts and behave unphysically \cite{stocker2022robust}. To achieve both requirements, we propose the use of two complementary methods to construct fragments, which allow models to learn different aspects of the PES of large systems. The first method follows a top-down approach, where fragments are constructed by ``cutting out'' spherical regions of the system of interest, which also includes solvent molecules in the condensed phase (Fig.~\ref{fig:fragment_construction}A) \cite{gastegger2017machine}. They are chosen as large as possible to sample important long-range effects, but still small enough such that reference energies and forces computed with quantum chemistry methods are accessible in a reasonable time. As our tests on poly-alanine systems demonstrate (see below), the top-down fragments we choose are sufficiently large for the systems studied in this work.

Since it is difficult to collect enough reference data for the large top-down fragments to train robust models, they are enriched by smaller fragments, for which data points for many different conformations can be collected. Starting from single atoms, molecules similar to local bonding patterns of the system of interest \cite{huang2020quantum} are systematically constructed by growing chemical graphs in a bottom-up fashion (Fig.~\ref{fig:fragment_construction}B). By limiting the size of these fragments, it is possible to sample many different conformations, allowing models to learn the effects of strong distortions in local structural patterns, which is key to preventing holes in the PES. As a result, the combination of bottom-up and top-down fragments enables learning accurate and robust MLFFs for large systems.

\subsection*{Poly-alanine systems}

\begin{figure*}[!htb]
    \centering
    \includegraphics[]{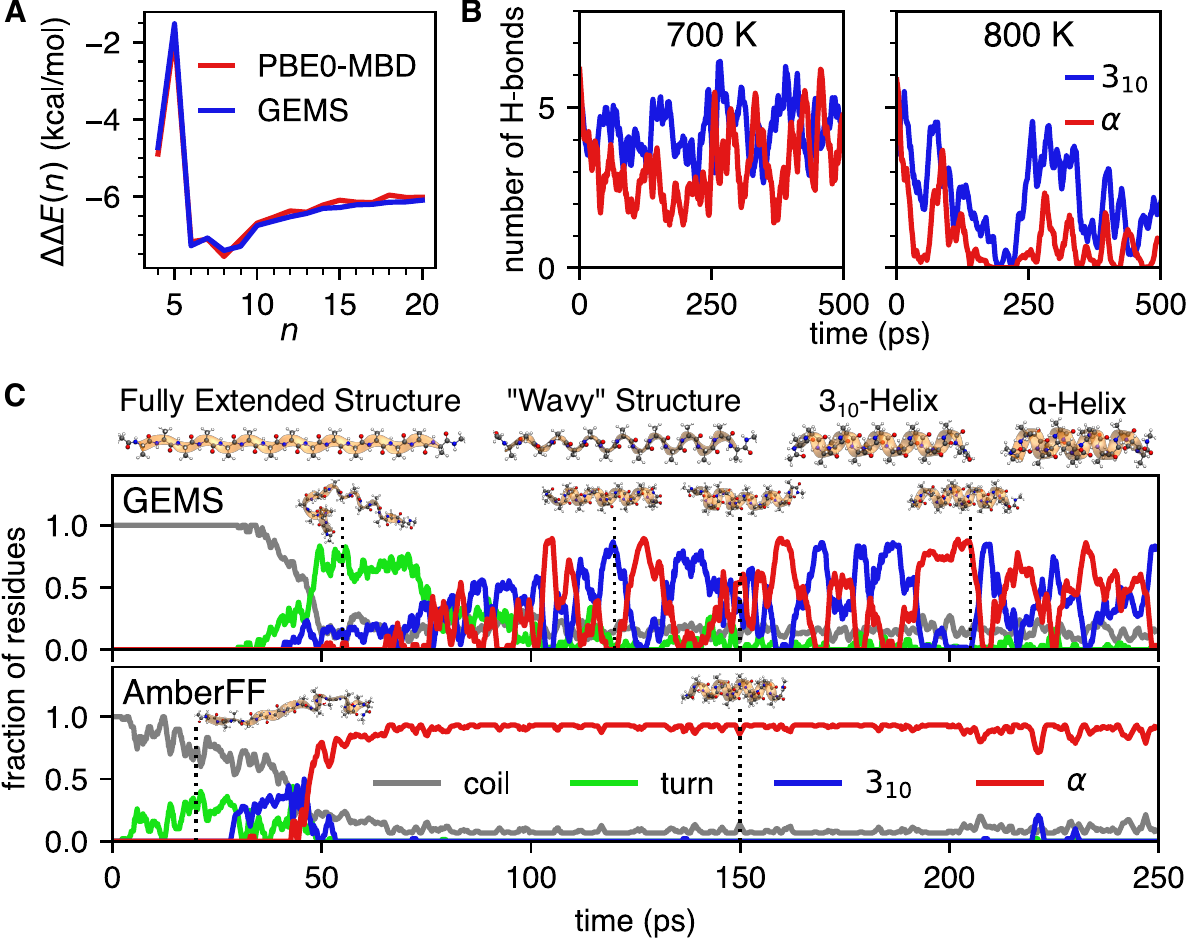}
    \caption{\textbf{Accurate simulations of poly-alanine systems with GEMS.}  (\textbf{A}) Relative stabilization of the $\alpha$-helical conformation of AceAla$_n$Lys~+~H$^+$ per added alanine residue. Shown here is the double difference $\Delta \Delta E(n) = \Delta E(n) - \Delta E(n-1)$, where $\Delta E(n) =  E_{\alpha}(n)-E_{\mathrm{FES}}(n)$ is the relative energy of the $\alpha$-helical conformation and the fully extended structure (FES) of AceAla$_n$Lys~+~H$^+$ in gas phase. The prediction of GEMS (blue) is compared to \textit{ab initio} reference data computed at the PBE0-MBD \cite{adamo1999toward,weigend2005balanced,tkatchenko2012accurate,ambrosetti2014} level of theory. (\textbf{B}) Number of $\alpha$- and $3_{10}$-helical H-bonds during MD simulations of helical AceAla$_{15}$Lys~+~H$^+$ in gas phase at 700~K and 800~K with GEMS. The sharp drop in the number of H-bonds in the dynamics at 800~K indicates the formation of a random coil (see Fig.~\ref{Sfig:hbonds_mlff}A for an extended version of this figure with a greater range of temperatures). (\textbf{C})
    Secondary structural motifs determined by STRIDE \cite{frishman1995knowledge} along typical folding trajectories of AceAla$_{15}$Nme. Dotted vertical lines indicate the temporal position of the shown snapshots. The trajectory computed with GEMS (top) folds via a distinct ``wavy'' intermediate (classified primarily as ``turn'') and settles into a dynamic equilibrium between $3_{10}$- and $\alpha$-helices. In contrast, the trajectory computed with the AmberFF (bottom) folds more directly and then stays primarily $\alpha$-helical (see Fig.~\ref{Sfig:acala15nme_additional_folding_trajectories} for an analysis of additional trajectories).}
    \label{fig:alanine_overview}
\end{figure*}

We apply GEMS to predict the properties and dynamics of several peptides consisting primarily of alanine. 
These are popular model systems for proteins and well studied both theoretically and experimentally. In addition, by limiting the number of residues, it is still possible to perform electronic-structure calculations for the full system. Thus, the predictions of an ML model trained only on fragment data can be directly compared to reference calculations, which allows to verify the ability of GEMS to reconstruct the properties of larger systems from the chemical knowledge extracted from smaller molecules.

As a first test case, we consider the cooperativity between hydrogen bonds in poly-alanine peptides capped with an N-terminal acetyl group and a protonated lysine residue at the C-terminus (AceAla$_n$Lys~+~H$^+$). In $\alpha$-helices, the local dipole moments of hydrogen bonds formed between backbone peptide groups are aligned, leading to a cooperative polarization effect \cite{park2000stabilization}. Thus, the relative stabilization energy of an $\alpha$-helix compared to a fully extended structure (FES) fluctuates non-trivially with helix length and is a challenging prediction task. We find that GEMS closely agrees with the reference \textit{ab initio} method, demonstrating that large-scale effects can be learned effectively from fragment data (Fig.~\ref{fig:alanine_overview}A).

Alanine-based peptides have a strong tendency to form helical structures. While short isolated helices are only marginally stable in solution, AceAla$_{15}$Lys~+~H$^+$ is known to form stable helices in gas phase. Experimental results suggest that AceAla$_{15}$Lys~+~H$^+$ remains helical up to temperatures of $\sim$725~K \cite{kohtani2004extreme}, allowing a direct comparison with theoretical predictions. By running GEMS simulations at different temperatures, we confirm that the peptide remains primarily helical up to 700~K, but forms a random coil at 800~K (see \href{https://youtu.be/QZIc3a4OjJk}{supplementary video~1} at \texttt{https://youtu.be/QZIc3a4OjJk}). An analysis of the formed hydrogen bonds reveals that the average number of $\alpha$-helical hydrogen bonds decreases with increasing temperature (see Fig.~\ref{Sfig:hbonds_mlff}A), while the number of $3_{10}$-helical hydrogen bonds remains almost constant until a sudden drop at 800~K (see Fig.~\ref{fig:alanine_overview}B), which agrees with results from \textit{ab initio} simulations \cite{tkatchenko2011unraveling}. Interestingly, the long-ranged interactions learned from top-down fragments seem to be crucial to reproduce the experimental results, as a model that was only trained on bottom-up fragments predicts reduced thermal stability (see Fig.~\ref{Sfig:hbonds_mlff}B).

To investigate whether there are fundamental differences between GEMS and dynamics simulations performed with conventional FFs, we study the room temperature (300~K) folding process of a pure poly-alanine peptide capped with an N-terminal acetyl group and a C-terminal N-methyl amide group (AceAla$_{15}$Nme) in gas phase. Starting from the FES, MD simulations with GEMS suggest that AceAla$_{15}$Nme has a strong tendency to form H-bonds between peptide groups of directly adjacent residues within the first $\sim$100~ps of dynamics. The formed arrangements exhibit a ``wavy'' structure and $\phi$~and~$\psi$ backbone dihedral angles of $\sim0^\circ$~and~$\sim0^\circ$, which lie in a sparsely populated region of the Ramachandran plot. These intermediates are typically short-lived with lifetimes of $\sim$25--50~ps, and fold readily into helical configurations via a characteristic twisting motion. There is still some controversy between theoretical and experimental results regarding the predominance of different helical conformations \cite{topol2001alpha}. We find that there may be cases where no single motif is preferred: Once a helix is formed, its structure fluctuates between pure $\alpha$- and $3_{10}$-helices, as well as hybrids of both helix types (see \href{https://youtu.be/ZuKW292DKKw}{supplementary
~video~2} at \texttt{https://youtu.be/ZuKW292DKKw} for a complete folding trajectory). A 10~ns trajectory of the helical state suggests a dynamical equilibrium with a $\sim$38\%/62\% mixture  of $\alpha$- and $3_{10}$-helices. In contrast, MD simulations with a conventional FF yield qualitatively different results, suggesting that a more rigid and primarily $\alpha$-helical configuration is formed from the FES without distinct structural intermediates (see Fig.~\ref{fig:alanine_overview}C).

As a final test for the accuracy of GEMS, we compare the predictions of the ML model to \textit{ab initio} data computed at the at the PBE0/def2-TZVPP+MBD \cite{adamo1999toward,weigend2005balanced,tkatchenko2012accurate,ambrosetti2014} level of theory. To this end, we use 1554 and 1000 AceAla$_{15}$Nme structures sampled from densely and sparsely populated regions (rare events) of the configurational space visited in 100 aggregated 250~ps MD trajectories (25~ns total) in the NVT ensemble at 300~K simulated with GEMS (see Section~\ref{Ssec:AlanineClusters} for details).
We find that predicted energies and forces are in good agreement with the reference values in both cases. For energies and forces, correlation coefficients are $R
^2=0.996$ and $R
^2=0.998$, respectively, and mean absolute errors (MAEs) are $0.450$~meV/atom and
 $36.704$~meV/\AA. Again, we find that the inclusion of top-down fragments during training is crucial for high accuracy, as prediction errors for a model trained only on bottom-up fragments are much larger (see Fig.~\ref{Sfig:correlations_ml_no_topdown}). In comparison, even though predictions with the conventional AmberFF \cite{Lindorff-Larsen2010AMBER99SB-ILDN} are remarkably accurate with correlation coefficients of $R
^2=0.928$ (for energy) and $R
^2=0.876$ (for forces), the MAEs are much larger at $2.274$~meV/atom and $329.328$~meV/\AA\ (distributions of predicted and reference energy values were shifted to have a mean of zero prior to computing MAEs in both cases, such that constant energy offsets between different methods do not influence the results). As a general trend, we observe that predictions with GEMS reproduce the reference across the whole range of values without the presence of a single outlier, whereas the AmberFF systematically under- and over-predicts small and large energy values, respectively (see Fig.~\ref{fig:overview}B). These findings show that GEMS gives accurate predictions even for rare configurations and the simulated MD trajectories are essentially \textit{ab initio} quality (see Figs.~\ref{Sfig:correlations_ml}~and~\ref{Sfig:correlations_ff} for a more detailed analysis of correlations within the different subsets of configurations). 

\subsection*{Crambin}

\begin{figure*}[!htb]
    \centering
    \includegraphics[width=\textwidth]{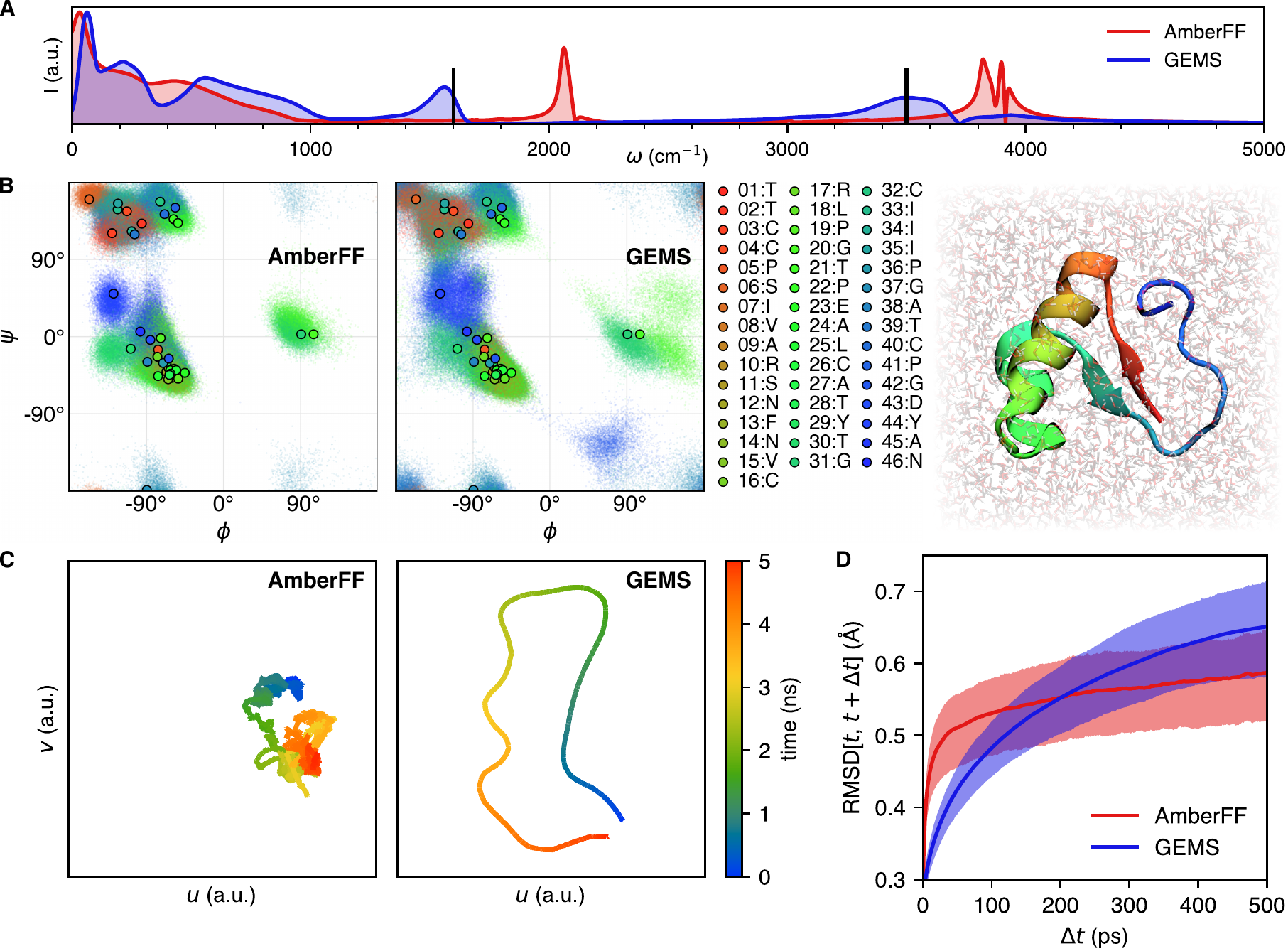}
    \caption{ \textbf{Analysis of dynamics simulations of crambin in aqueous solution.}
    (\textbf{A}) Power spectrum of crambin in water obtained from 125~ps of dynamics computed with the AmberFF and GEMS. In the GEMS spectrum, peaks associated with bending and stretching vibrations of water molecules are closer to experimentally expected values at around $\sim$1600~cm\textsuperscript{--1} and $\sim$3500~cm\textsuperscript{--1} \cite{ir_reference} (vertical black lines).
    (\textbf{B}) Ramachandran map for crambin (color-coded by residue number). The scatter shows the \mbox{($\phi$, $\psi$)}-dihedral angles sampled during an aggregated 10~ns of dynamics, points with black outline show values of the crystal structure \cite{bang2009role} for reference. Dynamics with GEMS (right) generally sample a broader distribution compared to AmberFF (left), indicating that the protein is more flexible. (\textbf{C}) Two-dimensional UMAP \cite{mcinnes2018umap} projection of the path through conformational space sampled during a 5~ns trajectory of crambin in aqueous solution. Compared to the trajectory computed with the AmberFF, dynamics with GEMS sample a wider distribution and are less likely to revisit previously visited regions of conformational space. (\textbf{D}) Distribution of root mean square deviations (RMSDs, excluding hydrogen atoms) between conformations sampled at times $t$ and $t+\Delta t$. Solid lines depict the mean, whereas the shaded region indicates the area between the 25th and 75th percentiles.  Dynamics with the AmberFF (red) show larger structural fluctuations on short time scales, whereas fluctuations on longer timescales are larger for dynamics computed with GEMS (blue).}
    \label{fig:crambin_overview}
\end{figure*}

GEMS enables accurate molecular simulations in the condensed phase. The 46-residue protein crambin in aqueous solution (25257 atoms) is chosen as a model system. Crambin contains 15 out of the 20 natural amino acids and forms common structural motifs such as $\beta$-sheets,  $\alpha$-helices, turns/loops, and disulfide bridges. 
To assess qualitative differences between simulations with a conventional FF (here, the AmberFF is chosen) and GEMS, we consider the power spectrum \cite{thomas2013computing} computed from 125~ps of dynamics at a temporal resolution of 2.5~fs (Fig.~\ref{fig:crambin_overview}A). The power spectrum is related to the internal motions of the system and reveals the dominant frequencies of molecular vibrations, which are influenced by the atomic structure and characteristic for the presence of certain functional groups. In comparison to the results obtained from the dynamics with a conventional FF, peaks in the power spectrum computed with GEMS are shifted towards lower wavenumbers and lie close to the frequency ranges expected from measured infrared spectra. For example, the dominant peaks above 1000~cm\textsuperscript{--1} correspond to bending and stretching vibrations of water molecules, which are experimentally expected at around $\sim$1600~cm\textsuperscript{--1} and $\sim$3500~cm\textsuperscript{--1}, respectively \cite{ir_reference}, which is consistent with the GEMS spectrum. In contrast, the corresponding peaks for the conventional FF are blue-shifted several hundreds of wavenumbers. Additionally, peaks in the 
GEMS spectrum are broader, indicating that the frequencies of characteristic vibrations are influenced stronger by intermolecular interactions, hence broadening their frequency range. Long-ranged interactions may particularly influence slow proteins motions, i.e.\ the low-frequency parts of the power spectrum, where notable differences between GEMS and AmberFF can be observed.

Similar to the results for AceAla$_{15}$Nme, we find that in comparison to simulations with the AmberFF, crambin is more flexible in GEMS simulations (Fig.~\ref{fig:overview}D). Although a straightforward quantitative comparison is not possible, qualitatively, the increased flexibility agrees more closely to structures modelled from nuclear magnetic resonance (NMR) spectroscopy measurements (see Fig.~\ref{Sfig:nmr_comparison}). The increased flexibility is also indicated by a Ramachandran map of the backbone dihedral angles of crambin (Fig.~\ref{fig:crambin_overview}B), which shows that a wider range of values is sampled in simulations with GEMS. This becomes even more apparent by projecting the trajectories into a low-dimensional space that allows a direct visualization of the path taken through conformational space (Fig.~\ref{fig:crambin_overview}C). However, a time-resolved analysis of the trajectories reveals that structural fluctuations with GEMS are only larger on timescales in excess of $\sim$200~ps (Fig.~\ref{fig:crambin_overview}D). On shorter time scales on the other hand, the trend is reversed. This suggests that there are qualitative differences between simulations with conventional FFs and GEMS on all timescales and simulations with \textit{ab initio} accuracy might be crucial to fully understand protein dynamics.

\section*{Discussion and conclusion}

\begin{figure*}[!htb]
    \centering
    \includegraphics{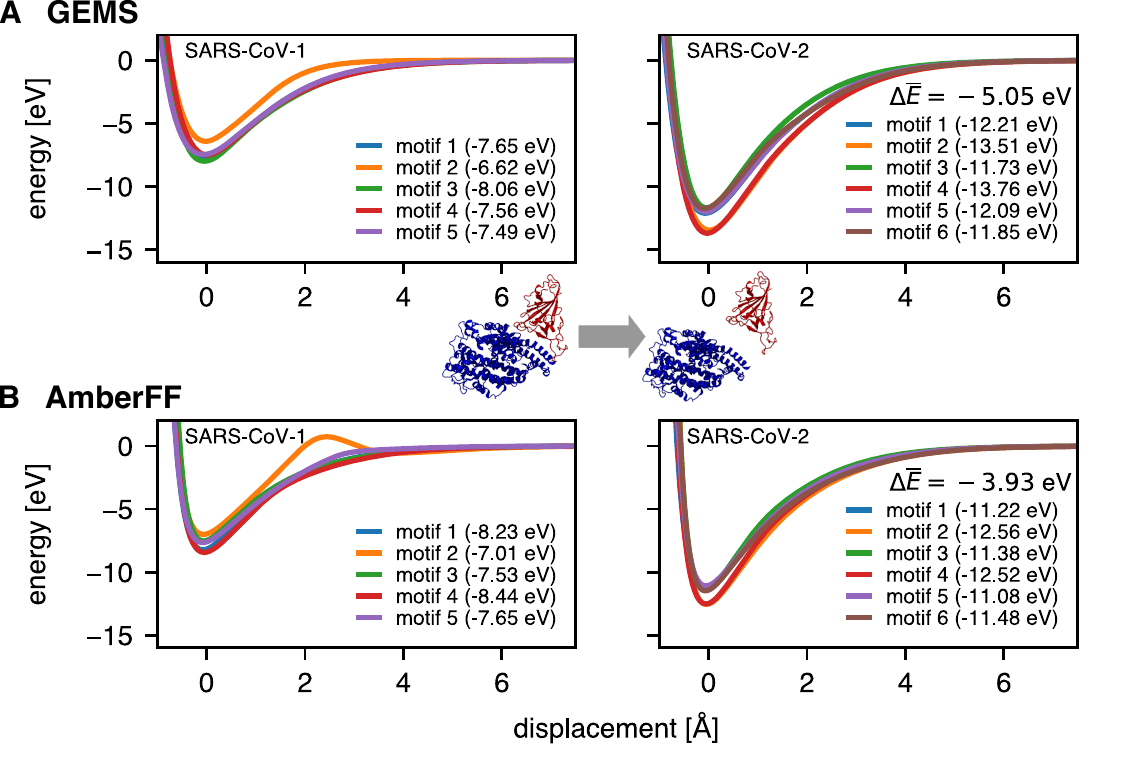}
    \caption{\textbf{Towards accurate quantum-mechanical protein--protein interactions: Gas-phase binding curves of the ACE2 (blue) and the receptor binding domain (RBD) of the SARS-CoV spike protein (red)}.  The ACE2 and RBD proteins are displaced along the line connecting their centers of mass relative to their equilibrium position in solution (computed with the AmberFF), keeping their internal structure fixed. Different binding motifs (taken from Ref.~ \citenum{delgado2021molecular}) are distinguished (values in brackets are the maximum well-depth for the corresponding motif). All energy values are referenced with respect to infinite separation of ACE2 and RBD. The displayed value $\Delta\overline{E}$ gives the difference in well-depth (averaged over all binding motifs) between SARS-CoV-2 and SARS-CoV-1. The $\Delta\Delta\overline{E}$ between AmberFF (\textbf{B}) and GEMS (\textbf{A}) is $-1.11$~eV.}
    \label{fig:cov12_scans}
\end{figure*}

%
Modeling quantum-mechanical properties of large molecules is an outstanding challenge and it holds promise for broad application in chemistry, biology, pharmacology, and medicine. We have developed a general framework for constructing MLFFs -- GEMS -- for large molecular systems such as proteins by learning from \textit{ab initio} reference data of  small(er) fragments without the need to perform electronic-structure calculations for a whole protein -- as the latter would constitute a computationally impractical task. The proposed divide-and-conquer strategy employing a library of $\sim$3 million DFT+MBD computations on fragments and using a machine learning model that incorporates physical constraints and long-range interactions allows to efficiently construct MLFFs that accurately reflect the quantum-mechanical energies and atomic forces in large molecules. An interesting insight of our \textit{ab initio} accurate simulations is that proteins are significantly more flexible than previously thought. These molecular fluctuations and associated low-frequency vibrations are expected to strongly contribute to dynamical processes such as protein folding, drug--protein binding, and allosteric regulation \cite{levy1982molecular}. 

While our work focuses exclusively on the study of peptides and proteins, the proposed framework can be applied to any atomic system too large to study with \textit{ab initio} methods. We find that even small poly-alanine peptides display qualitatively different dynamics when simulated with GEMS in comparison to dynamics with conventional FFs. For example, GEMS simulations suggest that the folding of AceAla$_{15}$Nme from the FES to a helical conformation occurs via short-lived intermediates characterized by hydrogen bonding between peptide groups of adjacent residues. Once a helix is formed, its structure fluctuates between $3_{10}$- and $\alpha$-helices in a dynamical equilibrium. This is in stark contrast to simulations with a conventional FF, where the peptide forms a rigid $\alpha$-helix without visiting a common intermediate. These results are reminiscent of the first MD study of a protein \cite{mccammon1977dynamics}, which showed that proteins are less rigid than previously thought \cite{phillips1981biomolecular}. The current findings, already alluded to above, indicate that proteins might be even more flexible, and our simulations of crambin suggest that the general trend observed for peptides in gas phase also holds for proteins in solution. In particular, crambin samples a larger conformational space in GEMS simulations and its backbone dihedral angles have broader distributions. Interestingly however, structural fluctuations on short time scales are reduced in comparison to simulations with a conventional FF. These observations show that there are qualitative differences in the dynamics of proteins when they are simulated with \textit{ab initio} quantum-mechanical accuracy. We conjecture that these variations in the dynamics might be crucial for the understanding of effects like allostery, or processes like enzyme catalysis and protein (mis)folding. For example, a description of the potential energy at \textit{ab initio} quality might be necessary to gain deeper insights into Levinthal's paradox \cite{levinthal1968there}.
A promising avenue for future work is to extend GEMS to larger systems and longer timescales, for example by distributing GEMS simulations over multiple accelerators, which requires non-trivial modifications to the way the MLFF is evaluated. Other possible extensions to GEMS include incorporating nuclear quantum effects, which were demonstrated to significantly change the dynamics of small molecules \cite{sauceda2021dynamical}. It is likely that similar effects can be observed for larger systems. \\

Let us discuss some limits of MLFFs when compared to classical MD simulations. Although MLFFs are orders of magnitude more computationally efficient than \textit{ab initio} calculations, their computational efficiency is lower than that of conventional FFs (as to be expected). For example, simulating a single timestep of NPT dynamics of crambin in aqueous solution on an NVIDIA A100 GPU with GEMS takes roughly $\sim$500~ms, whereas GROMACS \cite{hess2008gromacs} only requires $\sim$2~ms for a single time step with a conventional FF on similar hardware.
Consequently, at this moment, GEMS simulations are limited to shorter time scales. In addition, evaluating MLFFs usually requires increased memory compared to conventional FFs, limiting the maximum system size that can be simulated with GEMS. Nonetheless, GEMS allows to simulate several nanoseconds of dynamics for systems consisting of thousands of atoms with \textit{ab initio} accuracy (it may be possible to combine GEMS with other ML methods to achieve further speed-ups, e.g.\ by allowing larger time steps during the dynamics \cite{winkler2022super}).
Furthermore, GEMS like every other MLFF may lead to unphysical dynamics, if {\em not properly} trained (see e.g.\ Ref.~ \citenum{stocker2022robust} for a discussion of such phenomena). As a rule, MLFF simulations should therefore always be subjected to more scrutiny than results from mechanistic FFs. In particular, the resulting trajectories need to be carefully checked for unphysical bond breaking or formation, or otherwise unphysically distorted conformations, which are prevented in traditional FFs by construction. Nevertheless it should be emphasized again that compared to  simulations with a conventional FF, GEMS offers highly improved accuracy as well as enables to study chemical transformations such as the making and breaking of chemical bonds and proton transfer processes.

Another advantage of using accurate MLFFs is the availability of arbitrary derivatives -- including the potential to obtain alchemical derivatives \cite{saravanan2017alchemical,sheppard2010alchemical}. This may enable the {\em optimization} of accessible observables, such as docking/binding energies, with respect to local (nearly isosteric) mutations. 
In a more conventional approach, MLFFs can be used to describe the effects of mutations via thermodynamic integration as regularly performed with classical FFs \cite{blas2002theoretical,beierlein2011predicting,krepl2013effect}. Given the incorporation of non-locality in the present methodology, such analyses could naturally account for longer-range phenomena like (static) allosteric effects and the inherent non-additivity of interactions known to be relevant for the free energy of binding or stability~ \cite{dicera1998sitespecific}.
In a similar vein, this may allow to perform sensitivity analyses or engage explainability methods (see, e.g., Ref.~ \citenum{samek2021explaining}) to identify allosteric hotspots and networks, which have also been speculated to play an important role for the evolutionary aspects of proteins and the biomolecular machinery \cite{lockless1999evolutionarily,raman2016allostery}.
Again, we would like to stress that such studies and the above approaches are not limited to biomolecular systems. They may equally well be applied in materials design; for example, studying and optimizing point defects in solid state systems 
as relevant to the design of quantum materials.


Finally, we would like to highlight a promising application of GEMS to modeling protein-protein interactions. Fig.~\ref{fig:cov12_scans} shows the binding curves of the angiotensin converting enzyme~2 (ACE2) and the receptor binding domain (RBD) of the spike protein of SARS-CoV-1 and SARS-CoV-2 variants using either AmberFF or GEMS (in gas phase). Here, as expected from experimental evidence~ \cite{wrapp2020cryo} 
we observe a stronger binding of the SARS-CoV-2 both for the classical FF as well as GEMS. Interestingly, however, GEMS yields a substantially stronger binding by --1.1~eV. While the obtained binding energies do not account for solvation or entropic effects and cannot be directly compared to experimental binding affinities, this application provides evidence of the importance of \textit{ab initio} accuracy when studying interactions between complex biological systems. We therefore would like to stress the high promise of GEMS for enabling quantum-mechanical insight in broad application domains across enzyme and protein chemistry or heterogeneous materials. Although top-down fragments in this work are system-specific, in the future, they may be generated to cover a wider range of systems and enable GEMS simulations with a chemically transferable and size-extensive ``universal'' machine-learned force field.    


\section*{Materials and methods}

\subsection*{Construction of fragment data}
\paragraph*{Bottom-up fragments}

The construction of bottom-up fragments follows an approach similar in spirit to the one described by Huang et al.\ \cite{huang2020quantum}. 
Ignoring hydrogen atoms, increasingly large chemical graphs with the same local bonding patterns as the system of interest are constructed until a maximum number of heavy atoms is reached. This is achieved by starting from graphs consisting of a single heavy atom. Larger graphs are constructed by successively adding additional heavy atoms and pruning graphs which do not appear as substructures in the original system. Once the graphs are constructed, they are converted to bottom-up fragments by saturating all valencies with hydrogen atoms and generating the corresponding three-dimensional molecular structure (e.g.\ using Open Babel \cite{o2011open}). For all structures, multiple conformers are sampled using either MD simulations at high temperatures or normal mode sampling \cite{smith2017ani}. Here, we use the bottom-up fragments for proteins generated in earlier work~ \cite{unke2019solvated} (see Ref.~ \citenum{unke2019physnet} for details). Since all similar local structures are covered by the same graphs, the bottom-up fragments optimally exploit any structural redundancies, often resulting in a surprisingly small number of fragments. For example, just 2\,307 chemical graphs (with a maximum of eight heavy atoms) are sufficient to cover all local bonding patterns appearing in proteins consisting of the twenty natural amino acids, even when considering different protonation states and the possibility of disulfide bridges \cite{unke2019physnet}.

\paragraph*{Top-down fragments}
Starting from an MD snapshot of the system of interest (sampled from conventional MD simulations, see below and Section~\ref{Ssec:input_for_top_down_fragments} for more details), all atoms outside a spherical region around a central atom are deleted. The cutoff radius for selecting the spherical region should be chosen as large as possible, but still resulting in fragment sizes for which reference calculations are feasible (in this work, we choose 8~\AA). Then, any resulting dangling single bonds on heavy atoms are saturated with hydrogen atoms. Valencies situated on hydrogen atoms or corresponding to double bonds are eliminated by including the bonded atom in the original system (outside the cutoff). This process is repeated until all valencies are saturated \cite{gastegger2017machine}.  By choosing different central atoms, several (partially overlapping) top-down fragments can be constructed from a single configuration of the original system.

\subsection*{Density Functional Theory calculations}
Density Functional Theory (DFT) reference calculations were performed using the Psi4 software package \cite{parrish2017psi4,smith2018psi4numpy} at the PBE0/def2-TZVPP+MBD \cite{adamo1999toward,weigend2005balanced,tkatchenko2012accurate,ambrosetti2014} level of theory on Google Cloud Platform (GCP).
This level of theory has been found to be well-suited for modelling interactions of proteins in water \cite{distasio2014individual,stoehr2019quantum,schubert2015exploring}. Each fragment was run on an independent Docker container within a cloud compute engine virtual machine. We mostly used \emph{n2d-higmem-4} and \emph{n2-highmem-4} virtual machine instances with 4 cores, 32 GB RAM and 768 GB of disk space each, with some larger fragments being manually relaunched on higher-memory machines if they crashed with out-of-memory errors. Execution was parallelized on up to 20\,000 CPU cores. Calculations where shut down if they did not complete within 21 days, which was the case for a few outliers, but median execution time per fragment was $\approx 48$ hours.

\subsection*{Training the MLFF}
All MLFFs in this work use the recently proposed SpookyNet architecture (see Ref.~ \citenum{unke2021spookynet} for details). We use three different trained ML models in this work, one for the simulations of all poly-alanine systems, one for the simulation of crambin in aqueous solution, and one for the gas-phase ACE2/SARS~CoV-1/2~RBD binding curves shown in Fig.~\ref{fig:cov12_scans} (CoV model). The poly-alanine and CoV models use the recommended architectural hyperparameters of $T=6$ interaction modules and $F=128$ features \cite{unke2021spookynet}. Due to hardware limitations when performing MD simulations for thousands of atoms, the crambin model uses $T=3$ interaction modules and $F=64$ features to reduce memory requirements. All models use a short-range cutoff of $r_{\rm srcut} = 10~a_0$ ($\sim5.29$~\AA). The crambin and CoV models additionally employ a long-range cutoff of $r_{\rm lrcut} = 20~a_0$ ($\sim10.58$~\AA) for the computation of the analytical electrostatic and dispersion correction terms included in the SpookyNet energy prediction (to achieve sub-quadratic scaling with respect to the number of atoms). We follow the training protocol described in Ref.~ \citenum{unke2021spookynet} for fitting the parameters to reference energies, forces, and dipole moments, however, the mean squared loss function was replaced by the adaptive robust loss described in Ref.~ \citenum{barron2019general}. All models were trained on NVIDIA V100 GPUs on GCP using the same 2\,713\,986 bottom-up fragments, and 45\,948 (for the poly-alanine model), 5\,624 (for the crambin model), or 129\,942 (for the CoV model) top-down fragments. During training, structures were randomly drawn in equal amounts from bottom-up and top-down fragments, i.e.\ top-down fragments were oversampled to mitigate the imbalance in the numbers of bottom-up/top-down fragments.

\subsection*{MD simulations}
\paragraph*{Conventional FF}
All classical MD simulations have been performed with the GROMACS 2020.3 software package using NVIDIA V100 or A100 GPUs in a Kubernetes system on GCP. Throughout this work, we have employed the AMBER99SB-ILDN force field \cite{Lindorff-Larsen2010AMBER99SB-ILDN} for the conventional MD simulations. Standard amino acid definitions have been adapted to accommodate charged Lys~+~H$^+$ termini in accordance with the AMBER99SB-ILDN parametrizations where needed. 
In the MD simulations of ACE2/SARS CoV-2 RBD, the binding of the Zn$^{2+}$-cofactor in ACE2 has been described \textit{via} harmonic restraints to the experimentally-determined ligands in order to avoid potential shortcomings in the description of the metal--ligand interaction.
All solvated systems presented in this manuscript or used for sampling representative structures for generating top-down fragments were initially resolvated, optimized to a maximum atomic force of 1000~kJ/mol/nm, and equilibrated according to the protocol detailed in section~\ref{Ssec:md_simulations}. 
Simulations for studying non-equilibrium processes (i.e.\ the gas phase folding/unfolding of poly-alanine systems) have been started directly from optimized structures with velocities drawn from a Maxwell-Boltzmann distribution at twice the simulation temperature (such that the average kinetic energy during the simulation corresponds to the desired temperature \cite{Konermann2018pseudogp}). The gasphase simulations have thereby been realized in a pseudo-gasphase setting as proposed and validated in Ref.~ \citenum{Konermann2018pseudogp}.
All constant temperature MD simulations have been performed using temperature coupling \textit{via} stochastic velocity rescaling~ \cite{Bussi2007vrescale} and a Parrinello-Rahman barostat~ \cite{Parrinello1981barostat} has been used for NPT simulations. To speed up computations, standard MD simulations involved the commonly employed constraint of bonds involving hydrogen, while the power spectra reported in this work have been obtained from fully unconstrained simulations.
The starting structures of poly-alanine systems have been generated with the Avogadro software \cite{hanwell2012avogadro} and the initial structure of crambin has been taken from PDB entry 2FD7 \cite{bang2009role}, where the incorrectly described residues (SER11 and VAL15) have been remodeled using PyMOL \cite{delano2002pymol}. 
Our simulations of the ACE2/SARS CoV-1/2 RBD complex have been initiated from a set of representative conformations as identified in Ref.~ \citenum{delgado2021molecular} or pointwise mutations thereof. Currently available experimental results on the mutations present in the $\beta$-, $\gamma$-, $\delta$-, and $\epsilon$-variants of SARS CoV-2 do not indicate considerable structural changes to the spike RBD. After partial relaxation, simple pointwise mutations of the structural representatives obtained for the $\alpha$-variant can thus be assumed to represent viable starting points for MD simulations of the different variants.

\paragraph*{GEMS}
All MD simulations with the GEMS method were performed using the SchNetPack \cite{schuett2018schnetpack} MD toolbox with a timestep of 0.5~fs and without any bond constraints. Simulations for poly-alanine systems were performed on NVIDIA V100 GPUs on GCP, whereas crambin simulations were performed on NVIDIA A100 GPUs with 80GB. To mimic experimental conditions \cite{kohtani2004extreme}, the simulations of AceAla$_{15}$Lys~+~H$^+$ helix stability were performed in the NVE ensemble starting from an optimized structure with initial velocities drawn from a Maxwell-Boltzmann distribution at twice the simulation temperature as explained above. The folding simulations of AceAla$_{15}$Nme were performed in the NVT ensemble at 300~K starting from the optimized FES using the same method to assign initial velocities. Simulations of crambin in aqueous solution were performed in the NPT ensemble at a temperature of 300~K and a pressure of 1.01325~bar, starting from an optimized structure and initial velocities drawn from a Maxwell-Boltzmann distribution according to the simulation temperature (the first 1~ns of dynamics was discarded to allow the system to equilibrate). Constant temperature and/or pressure simulations use the Nos\'e-Hoover chain thermostat/barostat
 \cite{tobias1993molecular,martyna1996explicit} implemented in SchNetPack using a chain length of 3.

\section*{Acknowledgments}
We thank Michael Brenner for insightful comments.
OTU acknowledges funding from the Swiss National Science Foundation (Grant No. P2BSP2\_188147). MG works at the BASLEARN -- TU Berlin/BASF Joint Lab for Machine Learning, co-financed by TU Berlin and BASF SE. MS and AT acknowledge financial support from the Fond National de la Recherche Luxembourg (FNR) under AFR PhD grant ``CNDTEC (11274975)'' as well as from the European Research Council \textit{via} ERC Consolidator Grant ``BeStMo (725291)''.
This work was supported in part by the German Ministry for Education and Research under Grant Nos. 01IS14013A-E, 01GQ1115, 01GQ0850, 01IS18025A, 031L0207D, and 01IS18037A. KRM was partly supported by the Institute of Information \& Communications Technology Planning \& Evaluation (IITP) grants funded by the Korea government(MSIT) (No. 2019-0-00079, Artificial Intelligence Graduate School Program, Korea University and No. 2022-0-00984, Development of Artificial Intelligence Technology for Personalized Plug-and-Play Explanation and Verification of Explanation).
Correspondence to OTU, AT and KRM. 
There are no competing interests to declare. 
DFT reference data for training the models and scripts for running MD simulations with GEMS will be made available upon publication.

\bibliography{references}

\bibliographystyle{unsrtnat}


\makeatletter\@input{xsupplement.tex}\makeatother
\end{document}



\maketitle



\section{Background}
\label{sec:background}

Conventional force fields (FFs) allow to study large systems, e.g.\ entire viruses \cite{freddolino2006molecular,zhao2013mature,zimmerman2020citizen}, in atomic detail. They achieve this remarkable efficieny by modeling chemical interactions as a sum over simple empirical terms \cite{jones1924determination,gonzalez2011force,unke2020high}. However, while very efficient, their accuracy is limited \cite{vitalini2015dynamic} and they typically cannot describe chemical reactions. Although there are various efforts to increase the accuracy of classical FFs, for example by including polarization effects \cite{halgren2001polarizable,warshel2007polarizable} and sophisticated models for anisotropic charge distributions \cite{rasmussen2007force,darley2008beyond,kandathil2013accuracy,cardamone2014multipolar,unke2017minimal}, or by developing reactive FFs \cite{warshel1980empirical,van2001reaxff}, they are clearly much faster to evaluate but typically cannot compete with the accuracy of machine learned force fields (MLFFs) \cite{behler2007generalized,popelier2015qctff,smith2017ani,behler2017first,behler2021four,gdml,sgdml}. Machine learning (ML) methods ``learn the rules'' of quantum mechanics \cite{QML-Book} and their representation from data, allowing to skip computationally expensive \textit{ab initio} simulations. Beyond FF construction, there are several other applications of ML to quantum chemistry (QC). One of the earliest uses of ML in QC was the exploration of chemical space \cite{rupp2012fast,montavon2013machine,hansen2013assessment,hansen2015machine}. However, ML can also be used to accelerate studies that typically rely on MD simulations or other dynamical equations \cite{boninsegna2018sparse}. For example, it can be used to directly sample equilibrium distributions \cite{noe2019boltzmann,kohler2020equivariant} or rare events \cite{zhang2019targeted}, or directly predict reaction rates \cite{koner2019exhaustive}. Further, ML is used for predicting protein structure \cite{senior2020improved,jumper2021highly,tunyasuvunakool2021highly}, solving the Schr\"odinger \cite{carleo2017solving,pfau2020ab,hermann2020deep}, predicting wave functions \cite{schutt2019unifying,gastegger2020deep,unke2021se}, modelling solvated systems \cite{gastegger2021machine}, generating molecules and solving inverse design problems \cite{popova2018deep,popova2019molecularrnn,gebauer2018generating,gebauer2019symmetry,hoffmann2019generating,winter2019efficient,gebauer2022inverse}, and even for planning chemical syntheses \cite{strieth2020machine}.

For a more detailed overview of the use of ML in molecular and material science, refer to Refs.~\citenum{jain2013commentary,fischer2006predicting,tabor2018accelerating,butler2018machine,de2017use,hart2021machine}, for an overview of applications in molecular simulations, refer to Ref.~\citenum{noe2020machinemolsim}, and for reviews on the exploration of chemical space, refer to Refs.~\citenum{von2020exploring}~and~\citenum{huang2021ab}, furthermore  general reviews can be found in Refs.~\citenum{QML-Book,strieth2020machine,von2020retrospective,alexnatcoms2020,keith2021combining,glielmo2021unsupervised,meuwly2021machine,westermayr2021machine}.

\section{MD simulations --- Equilibration and detailed setup}
\label{sec:md_simulations}
After initial preparation (resolving doubly- or ill-defined residues and atom type definitions present in the original files from the respective sources specified in the main manuscript), classical molecular dynamics (MD) simulations of solvated systems were initialized by resolvating the systems in cubic simulation boxes with a minimum protein-to-box distance of 1.6~nm. Unless explicitly specified otherwise, simulation cells were solvated in TIP3P water with physiological concentrations of NaCl with excess Na\textsuperscript{+}- or Cl\textsuperscript{--}-ions to neutralize the simulation box where needed. The solvated structures were subsequently optimized to a maximum atomic force of 1'000~kJ/mol/nm and equilibrated in a four-step procedure consisting of (a time step of 2~fs was used in all cases):
\begin{itemize}[topsep=1mm,itemsep=0mm]
    \item[1)] a short NVT-simulation of 50'000 steps (simulation time 100~ps)
    \item[2)] NPT simulation (Berendsen barostat) of 50'000 steps (100~ps)
    \item[3)] NPT simulation (Parrinello-Rahman barostat~\cite{Parrinello1981barostat}) of 100'000 steps (200~ps) with fully constrained bonds
    \item[4)] and 100'000 steps (200~ps) with constraints on all bonds involving hydrogen.
\end{itemize}
In all equilibration runs a constant temperature thermostat with stochastic velocity rescaling~\cite{Bussi2007vrescale} set to the final simulation temperature was employed. Throughout all steps, the AMBER99SB-ILDN force field\cite{Lindorff-Larsen2010AMBER99SB-ILDN} was used.
        
For AcAla\textsubscript{15}Lys-chains involving the charged LysH\textsuperscript{+} terminus, topology and AMBER definitions have been adapted accordingly using the default AMBER99SB-ILDN parametrization. 




For the gasphase AcAla$_{15}$Lys+H$^{+}$, we adopted pseudo-gasphase settings as detailed in Ref.~\citenum{Konermann2018pseudogp} using maximal unit cells while disabling particle-mesh Ewald electrostatics. The constant temperature (pseudo-)gasphase simulations were prepared by structure optimization (maximum atomic force of 1'000~kJ/mol/nm). The reported simulations were then run in an NVT ensemble initialized with velocities randomly drawn from a Maxwell-Boltzmann distribution correponding to twice the simulation temperature.

\section{Sampling structures for top-down fragmentation}
\label{sec:input_for_top_down_fragments}
To train a model that can be used to simulate trajectories of a particular system of interest, we want to train it on a diverse set of top-down fragments representative of a variety of conformations. The general strategy is to cluster configurations that occur in classical MD simulations, select some representatives for each cluster, and then decompose the whole 
configurations into spherical regions that are small enough to run DFT calculations
on them (top-down fragments). Different systems have different characteristics when it comes to the possible
conformations: 
\begin{itemize}
   \item The poly-alanine systems unfold and thus show a lot of variation, but the overall system is comparatively small (contains few atoms).
 
  \item Crambin in aqueous solution contains many different atoms, but due to presence of three disulfide bridges, the protein itself shows only minimal variations, mainly determined by the states of the disulfide bridges, so clustering is straightforward.
 
\end{itemize}

\subsection*{Poly-alanine systems}
In our classical MD simulation of AceAla$_{15}$Nme and AceAla$_{n}$Lys~+~H$^{+}$ in solution at temperatures 280K, 300K, and 310K, (2 $\mu$s each)
the poly-alanine chain did not keep a helix structure, but assumed almost arbitrary
conformations. So we cannot define different well defined clusters, but
we can still use a clustering algorithm to find a diverse and representative 
sample of the configurations seen during the trajectories. \\
We used affinity propagation \cite{frey2007clustering}
as the clustering algorithm. This algorithm takes as input a matrix specifying
``similarities'' between two objects, and a ``preference'' that specifies 
the cost of adding a new cluster (which is balanced with the gain in similarity 
obtained by switching nearby objects to the new cluster). The output is a set
of clusters with one object in each cluster designated as the representative
of this cluster. 
The number of clusters is controlled by the relation between similarities and the
preference; default choices for the preference include the median similarity and the
lowest similarity, but in general the preference can be tuned to produce clusters at 
the desired granularity.\\
To compare two configurations of atoms, we move 
them so that the center of mass is at the origin, and then use the rotation that
gives the minimal mean square distance between the atoms. The similarity is then
the negative sum of the square distances.\\
We set the preference to -50 compared to a median similarity between -14 and -31 for
the six trajectories, this gave 240 cluster for AceAla$_{n}$Lys~+~H$^{+}$ molecule, 
and 266 cluster for the AceAla$_{15}$Nme molecule. 

\subsection*{Crambin}
Initial structure were taken from PDB entry \href{https://www.rcsb.org/structure/2fd7}{2FD7}. The incorrect residues SER11 and VAL15 have been remodeled using PyMOL. 

Crambin has 3 disulfide bridges at atoms (NCCS:SCCN) 
\begin{itemize}
  \item 31-33-35-38:561-558-556-554
  \item 41-43-45-48:449-446-444-442
  \item 220-222-224-227:373-370-368-366
\end{itemize}
These disulfide bridges have two stable positions, in two MD simulations over 
5 $\mu$s each we observed the first and the third disulfide bridge to flip between
stable positions (measured as dihedral angle of the N-C-C-S configurations). 
Together with a twist at an Arginine residue, these explained almost all the 
variations seen. We computed 22 clusters and made sure all observed variations were represented.

\section{Comparison to ground truth for AcAla\textsubscript{15}NME trajectories}
\label{sec:AlanineClusters}
To check the accuracy of our GEMS simulation, we select samples from 100 trajectories of 2500 steps
starting from a common stretched initial conformation. We subsample by only taking every 
second time step, which leaves 125,000 conformations. We use affinity propagation (see Section~\ref{sec:input_for_top_down_fragments}) to get representative samples. 
The similarity is the negative sum of square distances between corresponding non-hydrogen atoms,
after centering the molecules and applying an optimal rotation.
\\
However, using affinity propagation directly on these trajectories would have a large bias 
towards stable end conformations: Our trajectories contain stable end conformations
for roughly half of the time, so affinity propagation with the default settings would spend most
representatives on the stable conformations, largely ignoring the interesting folding part.
To reduce this bias, we use a preprocessing step that removes conformations that have a
small distance to an already selected point. (The threshold used was $9 \angstrom^2$ for the 
sum of the square distances, corresponding to $0.3 \angstrom$ per non-hydrogen atom for the RMSE.)
This reduces the stable tails of the trajectories, but if we do this only within the trajectories,
we get another bias around the common initial conformation. Computing pairwise distances
for the union of all trajectories would be computationally expensive, so we use an approximation:
We randomly mix all 100 trajectories and subdivide them into a partition of smaller subsets, 
and remove ``almost duplicates'' (as above) only inside each of the partitions. 
We then mix again and thin out again three more times. This removes
most of the ``almost duplicates'', and we arrive at 25,249 conformations
from the original 125,000 conformations. On these 25,249 we can then run the affinity propagation;
using preference$= -1500 \angstrom^2$ we arrive at 1554 representatives. 
Plotting the distance from the nearest representative (blue curve) over the 
trajectories now shows an even average distance: The red curve is a rolling average 
over 100 time points, it hovers around $0.4\angstrom$. The $x$-axis gives the time steps
in the simulation. After every 2500 steps = 250ps the next
trajectory starts, so in the image below there are data from the first two
trajectories.
In the blue curve, conformations that are selected as cluster representatives
can be seen as time 
points in which the blue curve touches the x-axis (since the distance to the closest representative
is then 0). We can see that there are regions that need more representatives (e.g. when folding happens),
and regions which only have occasional representatives (in the more stable end phase), 
but the average distance stays approximately constant.\\
%
\vskip0.5mm\noindent
\includegraphics[width=\columnwidth]{figures/avg_dist_repr.pdf}
\\
While this takes care of any obvious bias towards common or stable positions, we also add a 
list of 1000 conformations that are as far away as possible from all previously selected 
conformations. These can be thought of as untypical or unstable conformations, and we want to 
make sure that our model works for them as well as it does for the maybe more typical cluster
representatives.


\section{MD simulation code}
\begin{figure*}[!htbp]
    \centering
    \includegraphics[width=\textwidth, trim=0cm 4cm 0cm 4cm]{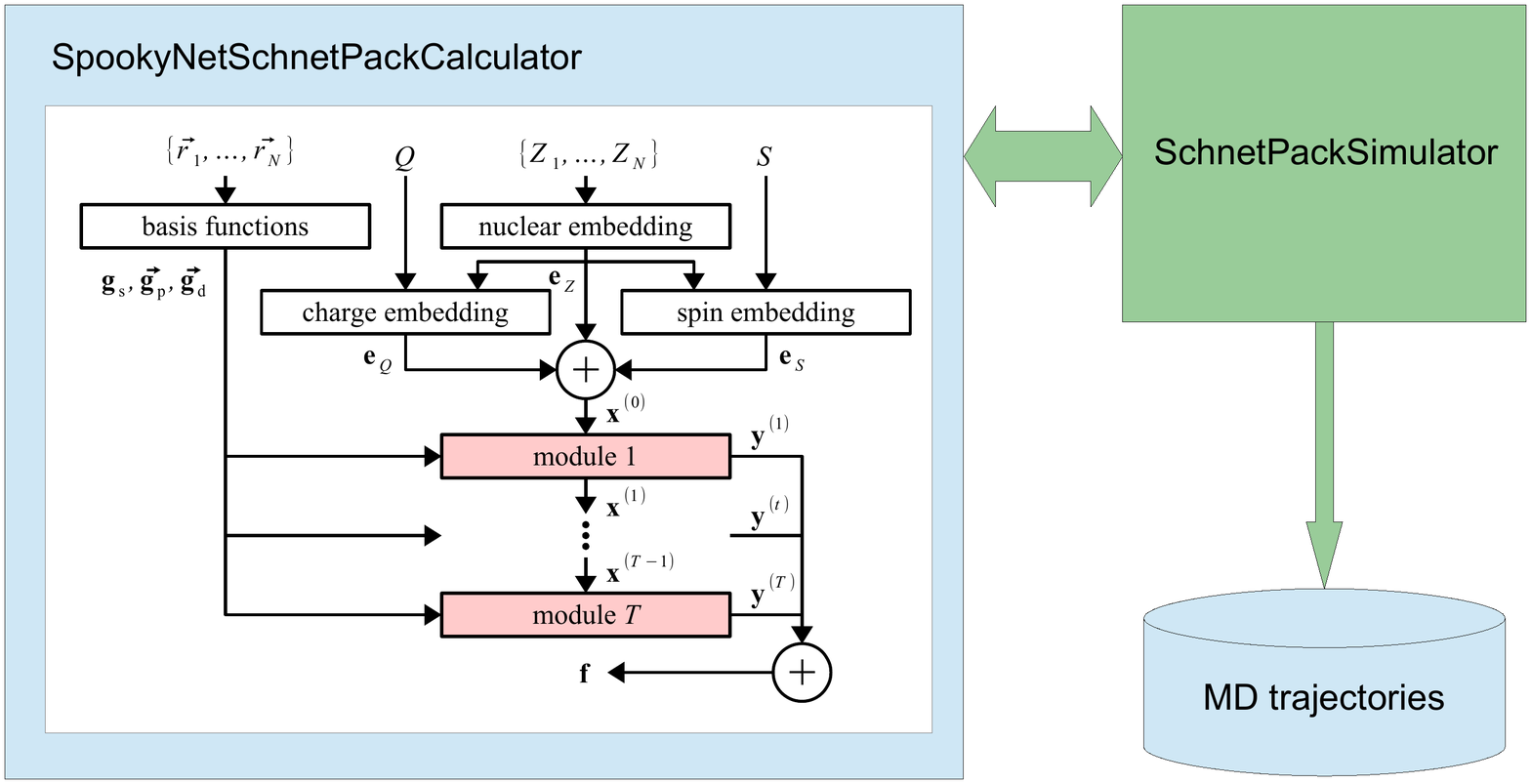}
    \caption{MD implementation: only the blue box (schnetpack.MDCalculator subclass using SpookyNet model to get forces predictions from atom positions and charges) needs to be implemented. SchNetPack Simulator takes care of running MD simulation, checkpointing and writing logs and trajectories to disk.}
    \label{fig:schnetpack}
\end{figure*}

The MD simulations are performed with the SchNetPack\cite{schuett2018schnetpack} toolbox providing an interface to the Atomic Simulation Environment \cite{larsen2017atomic} to run MD simulations with machine learning models. SchNetPack includes a fully functional MD suite, which can be used to perform efficient MD and PIMD simulations in different ensembles. The SpookyNet \cite{unke2021spookynet} model is used to implement \begin{verbatim}schnetpack.md.calculators.MDCalculator\end{verbatim}
interface from SchNetPack. See figure \ref{fig:schnetpack} for the schematic and corresponding papers for more details. Both SpookyNet and SchNetPack are written in PyTorch and thus can be used to run MD simulations directly on GPU increasing performance (and decreasing time per simulation step) by orders on magnitude compared to CPU-based models. 

\begin{figure*}[!htbp]
    \centering
    \includegraphics[scale=1.0]{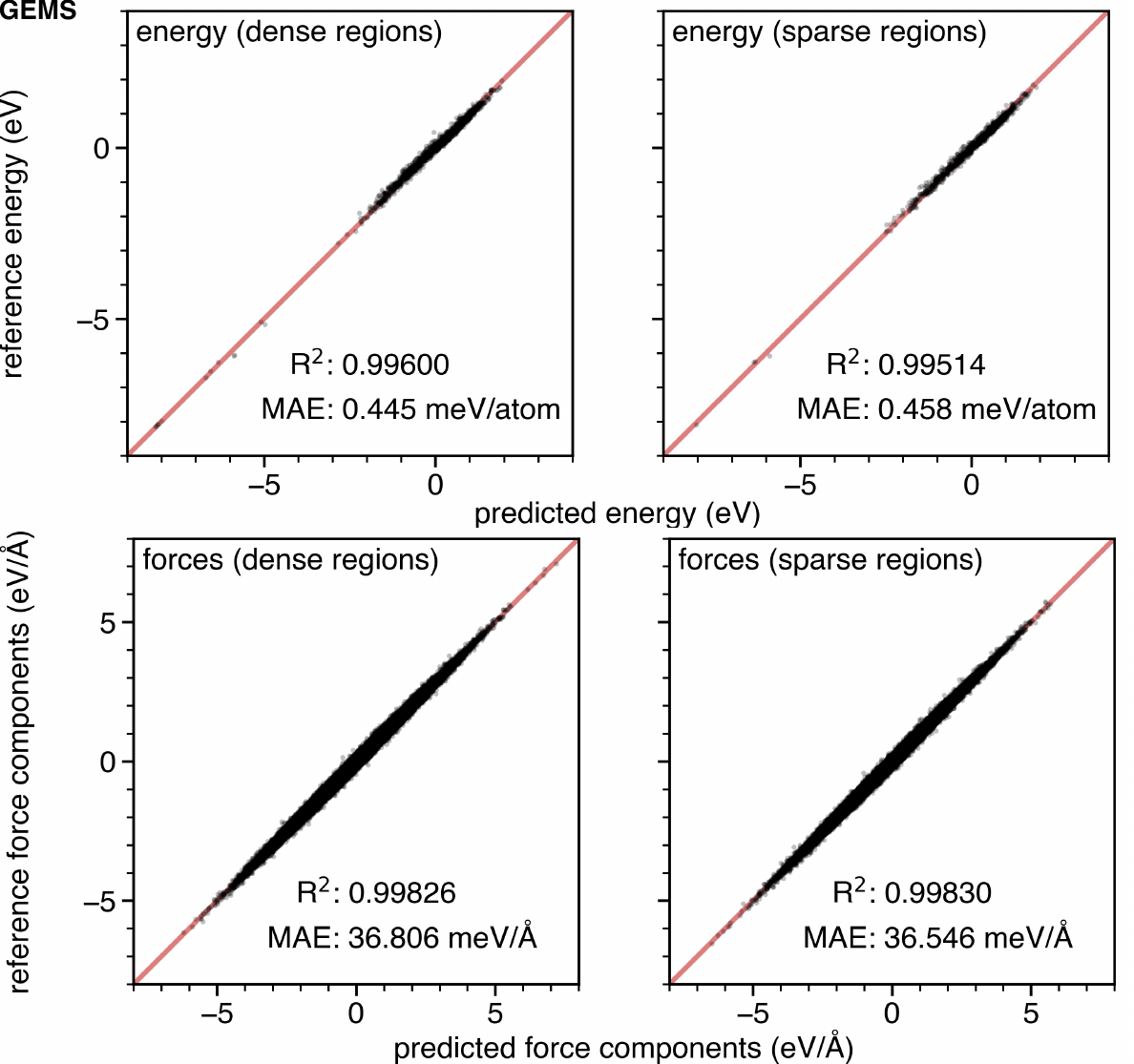}
    \caption{Correlation of predicted and \textit{ab initio} reference (ground truth) energies and forces for AceAla$_{15}$Nme conformations sampled from 100 aggregated 250~ps MD trajectories (25~ns total) in the NVT ensemble at 300~K simulated with GEMS. Conformations are sampled either from densely (1554 structures) or sparsely (1000 structures) populated regions of conformational space.}
    \label{fig:correlations_ml}
\end{figure*}

\begin{figure*}[!htbp]
    \centering
    \includegraphics[scale=1.0]{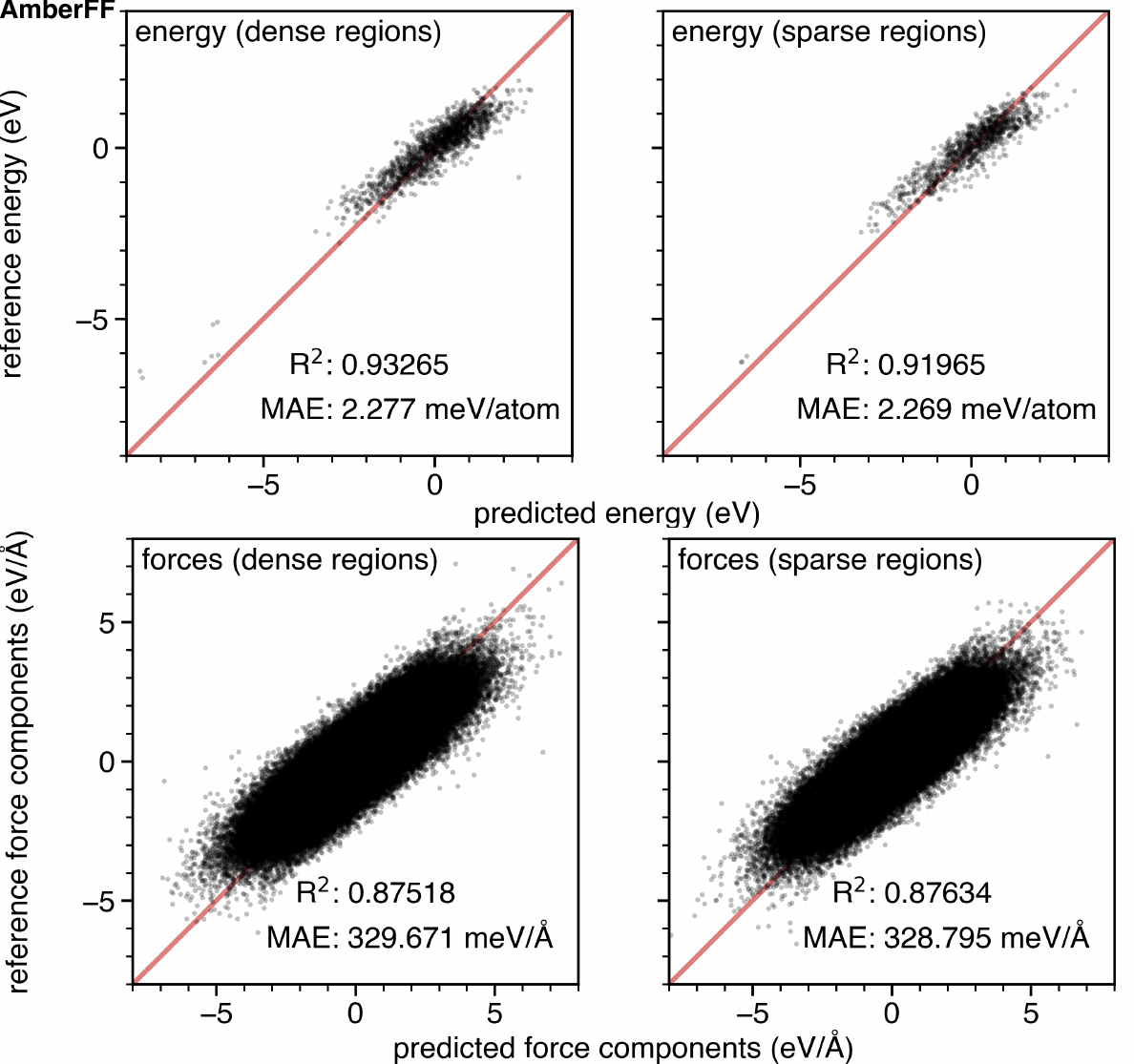}
    \caption{Same as Fig.~\ref{fig:correlations_ml}, but showing the correlation for predictions with the AmberFF.}
    \label{fig:correlations_ff}
\end{figure*}

\begin{figure*}[!htbp]
    \centering
    \includegraphics[scale=1.0]{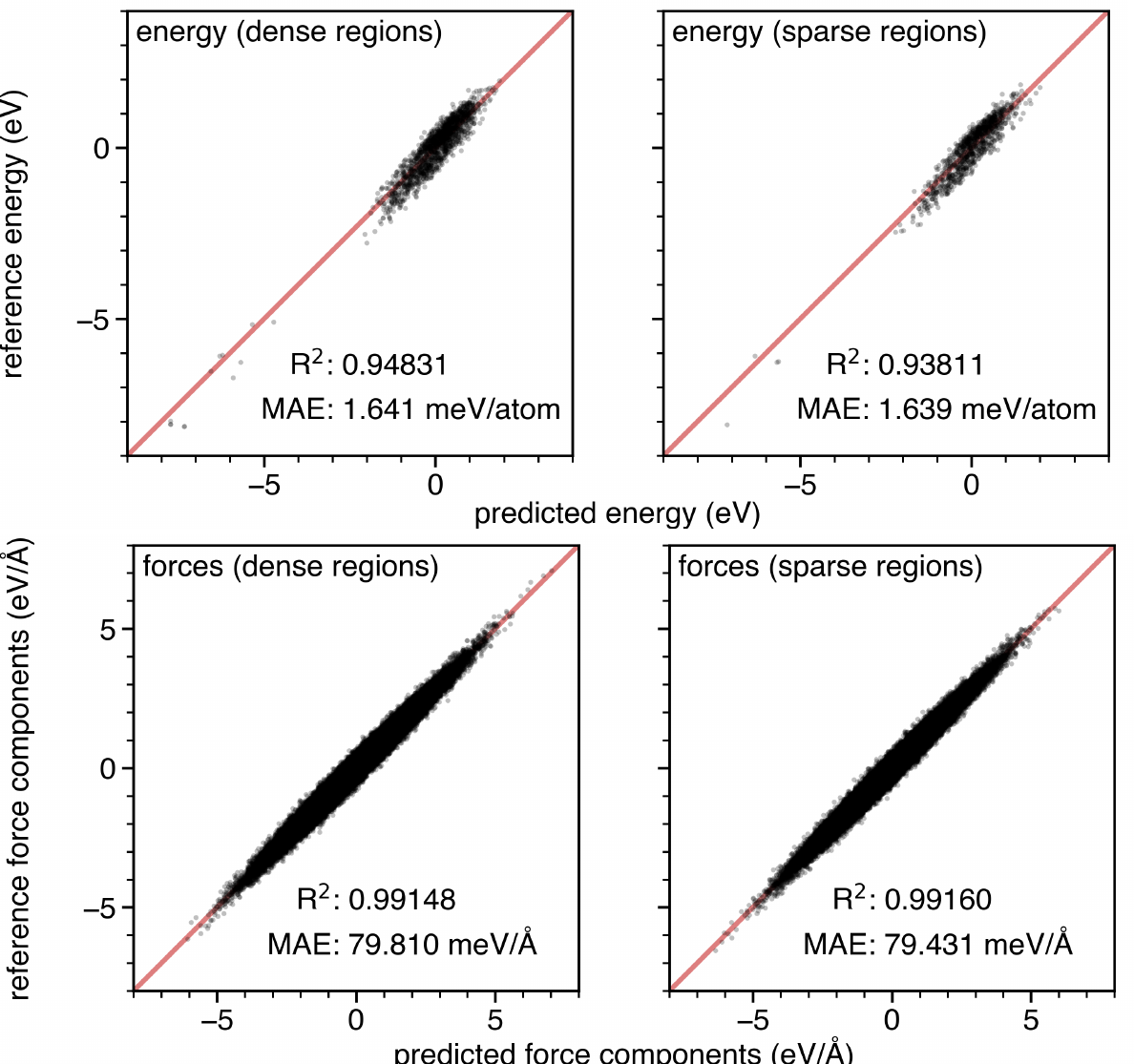}
    \caption{Same as Fig.~\ref{fig:correlations_ml}, but for a GEMS model trained without top-down fragments.}
    \label{fig:correlations_ml_no_topdown}
\end{figure*}

\begin{figure*}[!htbp]
    \centering
    \includegraphics[scale=1.0]{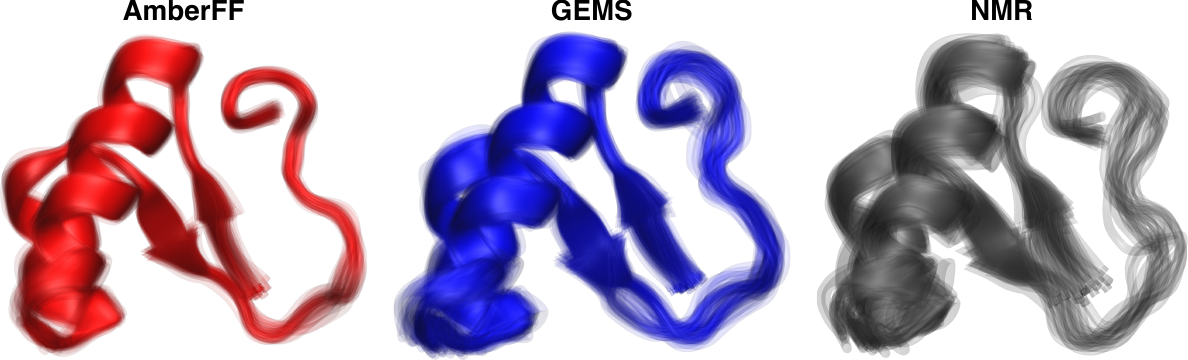}
    \caption{Overlay of representative conformations (obtained from cluster analysis) sampled during 10~ns of dynamics of crambin in aqueous solution. Simulations with GEMS (blue) lead to greater structural fluctuations compared to AmberFF (red), indicating that the protein is more flexible. For comparison, 20 low energy water refined structures of crambin in dodecylphosphocholine micelles based on NMR measurements are shown as well.\cite{ahn2006three} To allow a quantitative comparison, structures should be modelled with GEMS instead of a conventional FF when interpreting the NMR results.}
    \label{fig:nmr_comparison}
\end{figure*}

\begin{figure*}[!htbp]
    \centering
    \includegraphics[scale=1.0]{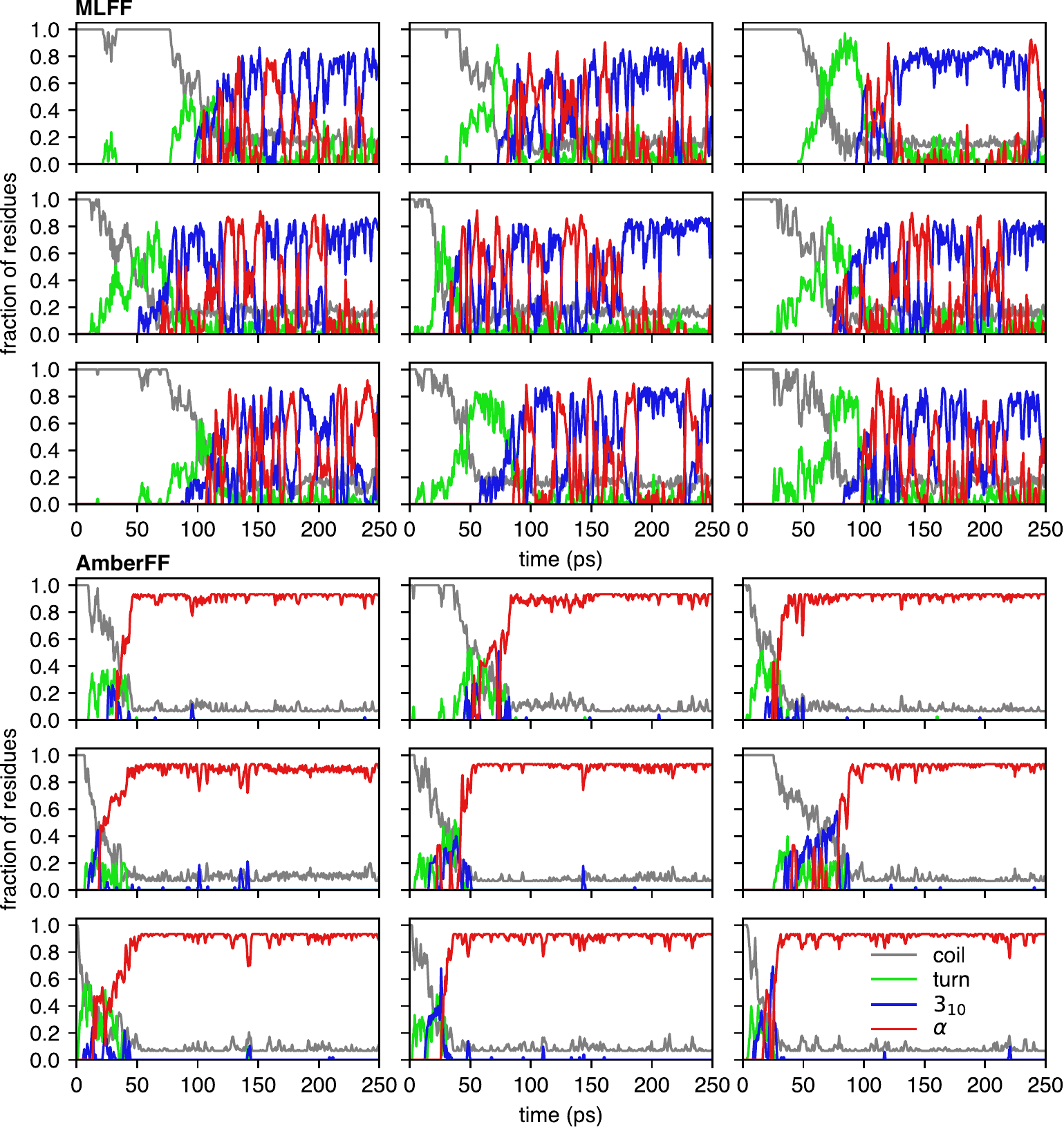}
    \caption{Secondary structural motifs (determined by STRIDE\cite{frishman1995knowledge}) along additional folding trajectories of AcAla$_{15}$NME.}
    \label{fig:acala15nme_additional_folding_trajectories}
\end{figure*}

\begin{figure*}[!htbp]
    \centering
    \includegraphics[scale=1.0]{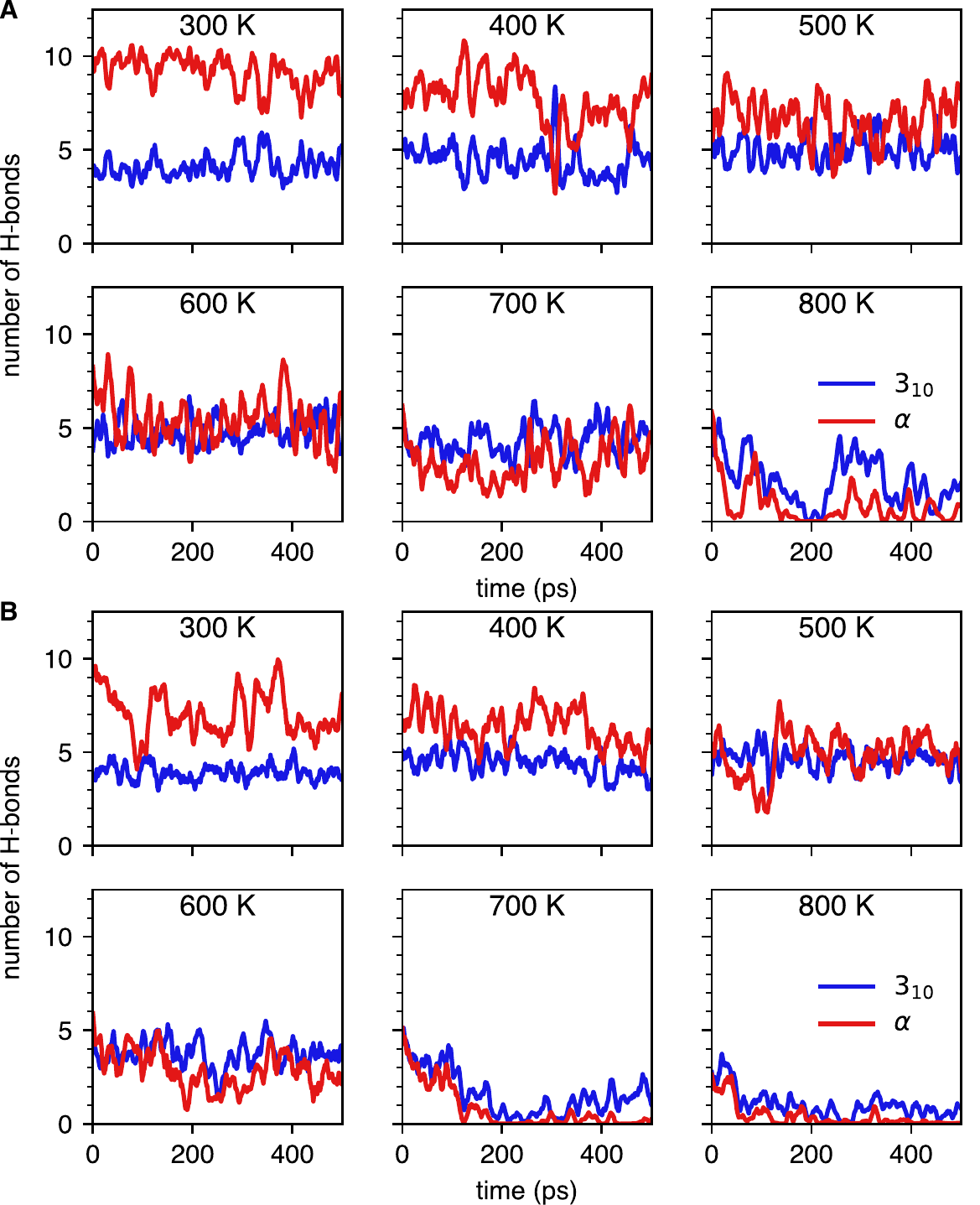}
    \caption{(\textbf{A}) Number of $\alpha$- and $3_{10}$-helical H-bonds during GEMS MD simulations of helical AceAla$_{15}$Lys~+~H$^+$ in gas phase at different temperatures. The sharp drop in the number of H-bonds in the dynamics at 800~K indicates the formation of a random coil. (\textbf{B}) Same as panel A, but for a model trained without top-down fragments. The number of H-bonds is lower on average for all temperatures and a random coil is formed at a lower temperature.}
    \label{fig:hbonds_mlff}
\end{figure*}

\bibliography{references}

\bibliographystyle{unsrtnat}

\makeatletter\@input{xmain.tex}\makeatother